\documentclass[aps,prd,onecolumn,groupedaddress,showpacs,nofootinbib,amssymb]{revtex4}
\usepackage[dvips]{graphicx}
\usepackage{amssymb}
\usepackage{amsmath}
\usepackage{graphicx,,color}
\usepackage{amsfonts}
\usepackage{bm}
\usepackage{cancel}
\usepackage{comment}
\newcommand\be{\begin{equation}}
\newcommand\ee{\end{equation}}

\allowdisplaybreaks[4]

\begin{document}

\title{$f(R)$-Gravity Generated Post-inflationary Eras and their Effect on Primordial Gravitational Waves}
\author{V.K. Oikonomou,$^{1}$}\email{v.k.oikonomou1979@gmail.com,voikonomou@auth.gr}
\affiliation{$^{1)}$Department of Physics, Aristotle University of
Thessaloniki, Thessaloniki 54124, Greece}

%$^{2)}$ Laboratory for Theoretical Cosmology, International Center
%of Gravity and Cosmos, Tomsk State University of Control Systems
%and Radioelectronics  (TUSUR), 634050 Tomsk, Russia

\tolerance=5000

\begin{abstract}
In this work we shall consider the effects of a geometrically
generated post-inflationary era on the energy spectrum of the
primordial gravitational waves. Specifically, we shall consider a
post-inflationary constant equation of state era, generated by the
synergistic effect of $f(R)$ gravity and of radiation and matter
perfect fluids. Two cases of interest shall be studied, one with
equation of state parameter $w=-1/3$, in which case the Universe
neither accelerates nor decelerates, and one with $w=0$ so an
early matter domination era. For the evaluation of the
inflationary observational indices which is relevant for the
calculation of the gravitational waves energy spectrum, we also
took into account the effects of the constant equation of state
parameter era, on the $e$-foldings number. In both the $w=-1/3$
and $w=0$ cases, the energy spectrum of the primordial
gravitational waves is amplified, but for the $w=0$ case, the
effect is stronger.
\end{abstract}

%PACS numbers: 04.50.Kd, 95.36.+x, 98.80.-k, 98.80.Cq

\maketitle

\section{Introduction}

Inflation is one of the most important theoretical proposals for
the early Universe
\cite{inflation1,inflation2,inflation3,inflation4}, since it
solves successfully all the shortcomings of the standard Big Bang
cosmology. However, to date there is no hint that inflation
occurred primordially, only constraints coming from Cosmic
Microwave Background (CMB) experiments, like the latest Planck
2018 collaboration \cite{Planck:2018vyg}. The occurrence of the
inflationary era can be confirmed in two ways, either by observing
the so-called $B$-modes in the CMB temperature fluctuations, or by
directly observing the stochastic primordial tensor perturbation
background in future experiments
\cite{Hild:2010id,Baker:2019nia,Smith:2019wny,Crowder:2005nr,Smith:2016jqs,Seto:2001qf,Kawamura:2020pcg,Bull:2018lat},
see also \cite{LISACosmologyWorkingGroup:2022jok} for the latest
updates for cosmological studies related to the LISA mission. So
far the CMB experiments have not yielded any hints of inflation,
however the stage 4 CMB experiments
\cite{CMB-S4:2016ple,SimonsObservatory:2019qwx} may observe the
$B$-modes of inflation directly on the CMB temperature
fluctuations. All these experiments are highly anticipated by the
theoretical physics, astrophysics and cosmology societies and will
provide sensational results. The stage 4 CMB experiments thus,
will directly verify the existence of the inflationary era via the
direct detection of the $B$-modes (curl modes) of inflation on the
CMB radiation temperature and polarization anisotropies. However,
the CMB merely probes modes with wavenumbers
$k<0.62$$\,$Mpc$^{-1}$, and thus primordial modes with larger
wavenumbers cannot be probed by the CMB temperature fluctuations,
since the CMB itself probes linear modes with wavelength $\lambda$
from $10$$\,$Mpc to $10^4$$\,$Mpc. Beyond that, non-linear
perturbation theory is needed, thus the CMB cannot reach the modes
which became subhorizon immediately after the inflationary era
during the reheating era. These modes which have wavelength below
$10$$\,$Mpc, are basically subhorizon modes, which exited the
Hubble horizon first during the inflationary era, and reentered
the horizon first, after inflation ended, thus became subhorizon
modes during the first stages of the reheating era and during the
subsequent radiation domination era. These tensor modes basically
constitute the primordial gravitational waves, and there exists a
vast literature on the theoretical aspects of the subject, see for
example,
\cite{Kamionkowski:2015yta,Denissenya:2018mqs,Turner:1993vb,Boyle:2005se,Zhang:2005nw,Schutz:2010xm,Sathyaprakash:2009xs,Caprini:2018mtu,
Arutyunov:2016kve,Kuroyanagi:2008ye,Clarke:2020bil,Kuroyanagi:2014nba,Nakayama:2009ce,Smith:2005mm,Giovannini:2008tm,
Liu:2015psa,Zhao:2013bba,Vagnozzi:2020gtf,Watanabe:2006qe,Kamionkowski:1993fg,Giare:2020vss,Kuroyanagi:2020sfw,Zhao:2006mm,
Nishizawa:2017nef,Arai:2017hxj,Bellini:2014fua,Nunes:2018zot,DAgostino:2019hvh,Mitra:2020vzq,Kuroyanagi:2011fy,Campeti:2020xwn,
Nishizawa:2014zra,Zhao:2006eb,Cheng:2021nyo,Nishizawa:2011eq,Chongchitnan:2006pe,Lasky:2015lej,Guzzetti:2016mkm,Ben-Dayan:2019gll,
Nakayama:2008wy,Capozziello:2017vdi,Capozziello:2008fn,Capozziello:2008rq,Cai:2021uup,Cai:2018dig,Odintsov:2021kup,Benetti:2021uea,Lin:2021vwc,Zhang:2021vak,Odintsov:2021urx,Pritchard:2004qp,Zhang:2005nv,Baskaran:2006qs,Oikonomou:2022xoq,Odintsov:2022cbm,Odintsov:2022sdk,Kawai:2017kqt,Odintsov:2022hxu,Gao:2019liu,}.
The era in which the primordial tensor modes became subhorizon,
namely the reheating era, is basically unknown to us, and the only
way to probe this era is via stochastic gravitational wave
experiments
\cite{Hild:2010id,Baker:2019nia,Smith:2019wny,Crowder:2005nr,Smith:2016jqs,Seto:2001qf,Kawamura:2020pcg,Bull:2018lat}.
The primordial gravitational waves will offer us insights, since
these are affected by numerous effects, like the matter content of
the Universe at horizon reentry, by the total equation of state of
the Universe at horizon reentry, by the number of massive and
massless particles at horizon reentry and so on. All these effects
will have direct impact on the stochastic gravitational wave
background, and thus will give us insights on these high energy
physics phenomena. Also supersymmetry breaking can also be probed
via the stochastic gravitational background, since if a huge
damping occurs for a specific frequency range, this means that
some particles obtained masses and supersymmetry broke at that
frequency range. This is a rather sensational feature. Finally,
let us note that the lowest frequency mode that is probed by the
CMB experiments corresponds to the tensor mode that reenters the
Hubble horizon today, and thus the maximum frequency that can be
reached by primordial gravitational waves spectrum is $f_{max}\sim
1$GHz \cite{Giovannini:2008tm}. The physics in the frequency range
$f_{BBN}<f<f_{max}$ (where $f_{BBN}$ is the frequency
corresponding to the Big Bang Nucleosynthesis)is unknown to us,
hence the primordial gravitational waves will actually reveal new
information related to several physical features of this unknown
to us era, not up to $f_{max}$ of course, but for frequencies with
maximum values several Hz.

Modified gravity in its various forms, can also describe
successfully inflation
\cite{reviews1,reviews2,reviews3,reviews4,reviews5,reviews6}. The
most profound modified gravity theory is $f(R)$ gravity
\cite{Nojiri:2003ft,Capozziello:2005ku,Hwang:2001pu,Cognola:2005de,Song:2006ej,Faulkner:2006ub,Olmo:2006eh,Sawicki:2007tf,Faraoni:2007yn,Carloni:2007yv,
Nojiri:2007as,Deruelle:2007pt,Appleby:2008tv,Dunsby:2010wg}, can
describe inflation and dark energy in a unified way, see for
example the pioneer work \cite{Nojiri:2003ft}. With regard to the
stochastic inflationary gravitational wave background, if a signal
is detected, this will really be intriguing theoretically. This is
because most ``traditional'' theories of standard inflation, like
$f(R)$ gravity and single scalar field theories, predict an
undetectable energy spectrum of primordial gravitational waves.
Thus if a signal is detected, one possibility of interpreting the
result is that these fundamental theories are excluded. Another
perspective is that it is possible that alternative evolution
scenarios took place post-inflationary. In this line of research,
in this article we aim to point out that if post-inflationary a
short lasting $f(R)$ gravity generated era with constant Equation
of State (EoS) parameter took place, then the predicted energy
spectrum of the primordial gravitational waves is significantly
enhanced. This post-inflationary era will be generated
synergistically by $f(R)$ gravity in the presence of matter and
radiation perfect fluids. We shall be interested to two distinct
constant EoS parameters, with $w=-1/3$ and $w=0$. The first
corresponds to a state that the Universe neither accelerates nor
decelerates, with $\ddot{a}=0$, while the second case corresponds
to an early matter domination era. As we show, in both cases, the
predicted energy spectrum of the primordial gravitational waves is
significantly enhanced, with the $w=0$ case producing a slightly
higher energy spectrum.

This paper is organized as follows: In section II, we present the
standard features of inflationary $f(R)$ gravity theories. Also we
discuss how a constant EoS post-inflationary era can be realized
by $f(R)$ gravity in the presence of perfect matter fluids.
Moreover, we show how the constant EoS short lasting
post-inflationary era may affect the duration of the inflationary
era. In section III we calculate the energy spectrum of the
primordial gravitational waves, for both the $w=-1/3$ and $w=0$
cases, and we show that the predicted signal is detectable by most
of the future gravitational wave experiments, if the reheating
temperature is large enough. Finally, the conclusions follow at
the end of the article.

\section{Inflation and Post-inflation Evolution with $f(R)$ Gravity}

As we mentioned in the introduction, our current perception of our
Universe indicates that the Universe went through four distinct
phases, the inflationary era, followed by the obscured reheating
era which consists partially from the radiation domination era,
followed by the matter and dark energy eras. To date we have
theoretical ideas of what drives inflation and all the
post-inflationary eras, but not clear-cut facts. Therefore, if the
inflationary and the dark energy eras are controlled by modified
gravity, there is no reason why modified gravity should not
control the rest of the cosmological eras, synergistically though
with the matter and radiation perfect fluids. In this line of
research, in this paper we shall assume that $f(R)$ gravity
actually controls entirely or synergistically all cosmological
eras, and the full form of it has the following approximate
behavior,
\[
f(R)=\left\{
\begin{array}{ccc}
  R+\frac{R^2}{6M^2}& R\sim R_I\, ,  \\
  F_w(R) & R\sim R_{PI}\ll R_I\, ,   \\
  F_{DE}(R) & R\sim R_0\ll R_{PI}\, , \\
\end{array}\right.
\]
where $R_I$ is the curvature scale of inflation, at the first
horizon crossing at the beginning of inflation when the comoving
mode $k$ leaves the Hubble horizon primordially. Also $R_{PI}$ is
the curvature scale during the early post-inflationary era, right
after inflation ends. Finally, $R_0$ is the curvature at present
day, so it is basically equal to the present day cosmological
constant. The functional forms of $F_w(R)$ and $F_{DE}(R)$ shall
be specified later on. Specifically, the function $F_w(R)$ will
generate a post-inflationary era with an EoS parameter $w$ which
is a constant and we shall specify it shortly. The function
$F_{DE}(R)$ is some conveniently chosen $f(R)$ gravity which can
produce a viable late-time phenomenology. Also, from the above
relation it is apparent that the inflationary era is controlled by
an $R^2$ gravity. Note that during the inflationary era, the
matter and radiation perfect fluids are neglected, but
post-inflationary they cannot be neglected. So both $F_w(R)$ and
$F_{DE}(R)$ synergistically with the matter and radiation perfect
fluids, control the evolution of the Universe.

Let us now use the metric $f(R)$ gravity formalism to find
$F_w(R)$. Consider the $f(R)$ gravity action in the presence of
perfect matter fluids,
\begin{equation}\label{action1dse}
\mathcal{S}=\frac{1}{2\kappa^2}\int
\mathrm{d}^4x\sqrt{-g}\,f(R)+\mathcal{S}_m,
\end{equation}
where $\kappa^2$ being equal to $\kappa^2=8\pi G=\frac{1}{M_p^2}$,
with $G$ is Newton's constant, and $M_p$ denotes the reduced
Planck mass. We shall use the metric formalism, so upon varying
the gravitational action with respect to the metric tensor, we
obtain the field equations,
\begin{equation}\label{eqnmotion}
f_R(R)R_{\mu \nu}(g)-\frac{1}{2}f(R)g_{\mu
\nu}-\nabla_{\mu}\nabla_{\nu}f_R(R)+g_{\mu \nu}\square
f_R(R)=0\kappa^2T_{\mu \nu}^{m}\, ,
\end{equation}
where $T_{\mu \nu}^m$ stands for the energy momentum tensor of the
matter and radiation perfect fluids, and also
$f_R=\frac{\mathrm{d}f}{\mathrm{d}R}$. Assuming that the geometric
background is that of a flat Friedmann-Robertson-Walker (FRW)
spacetime, the Friedmann equation is written,
\begin{equation}\label{frwf1}
-18\left (4H(t)^2\dot{H}(t)+H(t)\ddot{H}(t)\right)f_{RR}(R)+3
\left(H^2(t)+\dot{H}(t)
\right)f_{R}-\frac{f(R)}{2}+\kappa^2\left(\rho_m+\rho_r
\right)=0\, ,
\end{equation}
with $\rho_m$, $\rho_r$ denoting the energy density of the cold
dark matter and of the  radiation perfect fluids respectively. As
we mentioned, we shall assume that post-inflationary, for a short
period of time, the Universe has a constant EoS parameter $w$,
thus the total effective pressure $p_{eff}$ and the total energy
density $\rho_{eff}$, satisfy $p_{eff}=w\rho_{eff}$. Thus the
scale factor for this short post-inflationary period of time is,
\begin{equation}\label{scalefactorquasidesitter}
a(t)=a_{end}\,t^{\frac{2}{3(1+w)}}\, ,
\end{equation}
where $a_{end}$ is the scale factor at the end of the inflationary
era. Let us find which $f(R)$ gravity can generate the evolution
(\ref{scalefactorquasidesitter}) in the presence of matter and
radiation perfect fluids. We shall employ the reconstruction
techniques of Ref. \cite{Nojiri:2009kx} which are based on using
the $e$-foldings number as a dynamical variable,
\begin{equation}\label{efoldpoar}
e^{-N}=\frac{a_i}{a}\, ,
\end{equation}
with $a_i$ being some initial value of the scale factor. In terms
of the $e$-foldings number $N$, the Friedmann equation becomes,
\begin{equation}
\label{newfrw1}
-18\left [ 4H^3(N)H'(N)+H^2(N)(H')^2+H^3(N)H''(N)
\right ]f_{RR}(R)+3\left [H^2(N)+H(N)H'(N)
\right]f_R(R)-\frac{f(R)}{2}+\kappa^2\rho=0\, ,
\end{equation}
with $\rho=\rho_m+\rho_r$. Upon introducing the function,
$G(N)=H^2(N)$, we can write the Ricci scalar as follows,
\begin{equation}\label{riccinrelat}
R=3G'(N)+12G(N)\, .
\end{equation}
Thus by using a given scale factor, for example that of Ref.
(\ref{scalefactorquasidesitter}), we can solve everything with
respect to the cosmic time, find $t=t(a)$ and by using
(\ref{efoldpoar}) in conjunction with (\ref{riccinrelat}) we can
find the function $N(R)$. Thus by replacing in Eq.
(\ref{newfrw1}), we end up with the following second order
differential equation,
 \begin{equation}
\label{newfrw1modfrom} -9G(N(R))\left[ 4G'(N(R))+G''(N(R))
\right]f_{RR}(R) +\left[3G(N)+\frac{3}{2}G'(N(R))
\right]f_R(R)-\frac{f(R)}{2}+\kappa^2\left(\rho_m+\rho_r
\right)=0\, ,
\end{equation}
with $G'(N)=\mathrm{d}G(N)/\mathrm{d}N$ and
$G''(N)=\mathrm{d}^2G(N)/\mathrm{d}N^2$. Upon solving the above
equation, one can specify the function $f(R)$ which realizes the
given scale factor. Let us apply the method for the scale factor
(\ref{scalefactorquasidesitter}). For the scale factor
(\ref{scalefactorquasidesitter}) the function $G(N)$ takes the
form,
\begin{equation}\label{gnfunction}
G(N)=\frac{4 e^{-3 N (w+1)}}{9 (w+1)^2}\, ,
\end{equation}
with $a_r=1$ for convenience. Combining Eqs. (\ref{riccinrelat})
and (\ref{gnfunction}), we have,
\begin{equation}\label{efoldr}
N=\frac{\log \left(\frac{4 (1-3 w)}{3 R (w+1)^2}\right)}{3 (w+1)}.
\end{equation}
Taking the above into account, combined with the fact that
\begin{equation}\label{mattenrgydens}
\rho_{tot} =\sum_i\rho_{i0}a_0^{-3(1+w_i)}e^{-3N(R)(1+w_i)}\, ,
\end{equation}
the Friedmann equation (\ref{newfrw1modfrom}), takes the form,
\begin{align}
\label{bigdiffgeneral1} &a_1
R^2\frac{\mathrm{d}^2f(R)}{\mathrm{d}R^2}
+a_2R\frac{\mathrm{d}f(R)}{\mathrm{d}R}-\frac{f(R)}{2}+\sum_iS_{i}R^{
\frac{3(1+w_i) }{3(1+w)}}=0\, ,
\end{align}
with the index ``i'' taking the values $i=(r,m)$ with $i=r$
corresponding to radiation and $i=m$ corresponding to cold dark
matter perfect fluids. Also $a_1$ and $a_2$, $S_i$ and $A$ are
defined as follows,
\begin{equation}
\label{apara1a2} a_1=\frac{3(1+w)}{4-3(1+w)},\,\,\,
a_2=\frac{2-3(1+w)}{2(4-3(1+w))},\,\,\,S_i=\frac{\kappa^2\rho_{i0}a_0^{-3(1+w_i)}}{[3A(4-3(1+w))]^{\frac{3(1+w_i)}{3(1+w)}}},\,\,\,A=\frac{4}{3(w+1)}
\end{equation}
The solution of (\ref{bigdiffgeneral1}) is actually the $f(R)$
gravity which generates the post-inflationary evolution
(\ref{scalefactorquasidesitter}), which we denoted $F_w(R)$ and it
is,
\begin{equation}
\label{newsolutionsnoneulerssss} F_{w}(R)=\left
[\frac{c_2\rho_1}{\rho_2}-\frac{c_1\rho_1}{\rho_2(\rho_2-\rho_1+1)}\right]R^{\rho_2+1}
+\sum_i
\left[\frac{c_1S_i}{\rho_2(\delta_i+2+\rho_2-\rho_1)}\right]
R^{\delta_i+2+\rho_2}-\sum_iB_ic_2R^{\delta_i+\rho_2}+c_1R^{\rho_1}+c_2R^{\rho_2}\,
,
\end{equation}
with $c_1,c_2$ being integration constants, and also we introduced
$\delta_i$ and $B_i$,
\begin{equation}
\label{paramefgdd}
\delta_i=\frac{3(1+w_i)-23(1+w)}{3(1+w)}-\rho_2+2,\,\,\,B_i=\frac{S_i}{\rho_2\delta_i}
\, ,
\end{equation}
with $i=(r,m)$. Recall that we have assume that,
\[
f(R)=\left\{
\begin{array}{ccc}
  R+\frac{R^2}{6M^2}& R\sim R_I\, ,  \\
  F_w(R) & R\sim R_{PI}\ll R_I\, ,   \\
  F_{DE}(R) & R\sim R_0\ll R_{PI}\, , \\
\end{array}\right.
\]
so post-inflationary, the $f(R)$ gravity is given by $F_w(R)$
given in Eq. (\ref{newsolutionsnoneulerssss}). Regarding the
late-time era, the $f(R)$ gravity can be of power-law type such as
the one studied in Ref. \cite{Odintsov:2021kup}, which we will use
too in this work, and it is,
\begin{equation}\label{starobinsky}
F_{DE}(R)=-\gamma \Lambda
\Big{(}\frac{R}{3m_s^2}\Big{)}^{\delta}\, ,
\end{equation}
with $m_s$ in Eq. (\ref{starobinsky}) being defined as
$m_s^2=\frac{\kappa^2\rho_m^{(0)}}{3}$, and also $\rho_m^{(0)}$ is
the energy density of cold dark matter at present day. Moreover,
the parameter $\delta $ takes values in the interval $0<\delta
<1$, and $\gamma$ is a dimensionless parameter, while $\Lambda$ is
the present day cosmological constant. The model
(\ref{starobinsky}) generates a viable dark energy era, as it was
proven in detail in \cite{Odintsov:2021kup}. Thus during the
inflationary era the evolution is driven by the $R^2$ model, while
post-inflationary and for a short period, the evolution is
described by $F_w(R)$ given in Eq.
(\ref{newsolutionsnoneulerssss}), while at late times the
evolution is driven by $F_{DE}(R)$ given in Eq.
(\ref{starobinsky}). Such a description describes the evolution in
different patches, so an effective model which can potentially
describe in a unified way the different evolution patches we
described above is,
\begin{align}\label{mainfreffective}
& f(R)=e^{-\frac{\Lambda }{R}}\left(R+\frac{R^2}{6 M^2}
\right)+e^{-\frac{R}{\Lambda }}\tanh
\left(\frac{R-R_{PI}}{\Lambda} \right)F_{DE}(R)
+e^{-\frac{R}{\Lambda }}\tanh \left(\frac{R-R_0}{\Lambda}
\right)\left(F_w(R)-R-\frac{R^2}{6 M^2} \right)\, ,
\end{align}
where $R_{PI}$ is the curvature during the era with EoS parameter
$w$ and also $R_{0}$ stands for the curvature at present day. We
need to note that the above model is a theoretical approximation
of the full functional form of the $f(R)$ gravity, not an exact
solution of the Friedmann equation. Hence in the way given above,
one needs to have available the full functional form of the scale
factor, which is simply impossible to find. However, the different
patches of the $f(R)$ gravity can satisfy the Friedmann equation
(\ref{newfrw1}) or equivalently (\ref{frwf1}), given the scale
factor. For example, when $R\sim R_{PI}$, the effective model
(\ref{mainfreffective}) yields $f(R)\sim F_w(R)$ and this
approximate form satisfies exactly the Friedmann equations
(\ref{newfrw1}) or equivalently (\ref{frwf1}) for the scale factor
(\ref{scalefactorquasidesitter}). In fact the solution
(\ref{newsolutionsnoneulerssss}) is an exact solution of the
Friedmann equation (\ref{newfrw1}) for the scale factor being
(\ref{scalefactorquasidesitter}), in the presence of matter and
radiation fluids.

Of course the above model is an effective description which
realizes in a unified way the different evolutionary patches we
described previously, so many other models may describe such a
behavior. The model (\ref{mainfreffective}) is just an example,
out of many which may effective describe the different evolution
patches we described previously. In all the cases though, the
total $f(R)$ gravity must satisfy several theoretical constraints
and also the leading order form of the $f(R)$ gravity at present
time must respect the solar system constraints.

Regarding the theoretical constraints that must be satisfied by
any $f(R)$ gravity at all eras, these are, \cite{reviews1},
\begin{equation}\label{viabilitycriteria}
f'(R)>0\,, \,\,\,f''(R)>0\, ,
\end{equation}
for $R>R_{0}$, with $R_{0}$ being the curvature of the Universe at
present day. Hence, the model and every other model that can
describe the different patches of the $f(R)$ gravity quoted below
Eq. (\ref{paramefgdd}), must satisfy the viability criteria
(\ref{viabilitycriteria}) for all curvatures up to $R\sim R_I$
with $R_I\sim 12H_I^2$, and with $H_I$ being the inflationary
scale, which is $H_I\sim 10^{13}\,$GeV in the low scale
inflationary scenarios. Let us see whether the effective model
(\ref{mainfreffective}) satisfies the viability criteria
(\ref{viabilitycriteria}). For $R\sim R_{PI}$, so when the
Universe is during the post-inflationary $w$-era, we have $\tanh
\left(\frac{R-R_{PI}}{\Lambda} \right)\sim 0$, and in addition,
$\Lambda/R_{PI}\sim 0$ thus $e^{-\frac{\Lambda }{R}}\sim 1$, and
moreover $\tanh \left(\frac{R-R_0}{\Lambda} \right)\simeq 1$ since
$R_{PI}\gg R_0$, hence the $f(R)$ gravity is approximately
described by $f(R)\simeq F_w(R)$. We have checked these
constraints numerically for the form of $F_w(R)$ we found in Eq.
(\ref{newsolutionsnoneulerssss}) previously, and viability
constraints (\ref{viabilitycriteria}) are satisfied for both the
values of $w$ we shall use, namely $w=-1/3$ and $w=0$. Also it
must be noted that the free chosen constants $c_1$ and $c_2$ must
take negative values in order for the viability criteria
(\ref{viabilitycriteria}) to be satisfied. Furthermore when $R\sim
R_0$ at present day, then  $\tanh \left(\frac{R-R_0}{\Lambda}
\right)\simeq 0$, thus the $f(R)$ gravity at present day is at
leading order $f(R)=R+\frac{R^2}{6 M^2}+F_{DE}(R)$ with the
functional form of $F_{DE}(R)$ being quoted in Eq.
(\ref{starobinsky}). At late times, the viability criteria of the
model $f(R)=R+\frac{R^2}{6 M^2}+F_{DE}(R)$ have been investigated
in Ref. \cite{Odintsov:2020nwm} (see figure 5 of Ref.
\cite{Odintsov:2020nwm}) for both the late-time era and also for
the case that $R\sim 12 H_I^2$ with $H_I$ being the inflationary
scale. In the latter case we have $R\sim 12 H_I^2$ and $f(R)\sim
R+\frac{R^2}{6 M^2}$ so for $H_I\sim 10^{13}\,$GeV we have (see
also \cite{Odintsov:2020nwm} below Eq. (47)) $f''(R)\sim
2.15\times 10^{-28}\,$eV$^{-1}$ and $f'(R)\sim 3.59$. Thus the
viability conditions are satisfied, and also regarding the solar
system tests, the model at late times $f(R)=R+\frac{R^2}{6
M^2}+F_{DE}(R)$ is known to pass the solar system constraints, see
for example the review \cite{reviews1}.

Essential for the calculation of the primordial gravitational
waves energy spectrum, are the tensor-to-scalar ratio and the
tensor spectral index. In most cases in the literature,
post-inflationary the Universe is described by the radiation era
which is followed by the matter domination era. In these cases,
the modes which exited the horizon first during the beginning of
inflation, are affected in the standard way described in the
literature, via the $e$-foldings number which measures the
duration of the inflationary era, so usually $N\sim 60$. However,
even a short post-inflationary era with general EoS parameter $w$
different from that of radiation, will have a direct effect on the
$e$-foldings number. Specifically, the $e$-foldings number for a
primordial mode $k$ which exited the horizon during the first
horizon crossing at the beginning of inflation, is given by
\cite{Adshead:2010mc},
\begin{equation}\label{generalefoldingsnumber}
\frac{a_kH_k}{a_0H_0}=e^{-N}\frac{H_ka_{end}}{a_{reh}H_{reh}}\frac{H_{reh}a_{reh}}{a_{eq}H_{eq}}\frac{H_{eq}a_{eq}}{a_{0}H_{0}}\,
,
\end{equation}
where $a_k$ and $H_k$ are the scale factor and the Hubble rate at
the moment in which the primordial mode $k$ exits the horizon at
the beginning of the inflationary era, $a_{end}$ is the scale
factor at the end of inflation, $a_{reh}$ and $H_{reh}$ are the
scale factor and the Hubble rate at the end of the reheating era.
Also, $a_{eq}$ and $H_{eq}$ are the scale factor and the Hubble
rate at the matter-radiation equality, and $a_0$ and $H_0$ are the
scale factor and the Hubble rate at present day. Assuming that
post-inflationary, right after the end of the inflationary era,
the effective EoS parameter is constant $w$, we easily obtain
that,
\begin{equation}\label{aux1}
\ln \left(\frac{a_{end}H_{end}}{a_{reh}H_{reh}}
\right)=-\frac{1+3w}{6(1+w)}\ln
\left(\frac{\rho_{reh}}{\rho_{end}} \right)\, ,
\end{equation}
where $H_{end}$ is the Hubble rate at the end of inflation and
$\rho_{end}$ and $\rho_{reh}$ are the energy density of the
Universe at the end of the inflationary era and of the reheating
era respectively. Also for deriving the relation of Eq.
(\ref{aux1}) we assumed that at the time instances at the end of
inflation and at the end of the reheating era, the effective EoS
parameter is $w$. If post-inflationary the Universe is described
by a constant effective EoS parameter $w$ short era, followed by
standard radiation and matter domination eras, the $e$-foldings
number is easily found to be \cite{Adshead:2010mc},
\begin{equation}\label{efoldingsmainrelation}
N=56.12-\ln \left( \frac{k}{k_*}\right)+\frac{1}{3(1+w)}\ln \left(
\frac{2}{3}\right)+\ln \left(
\frac{\rho_k^{1/4}}{\rho_{end}^{1/4}}\right)+\frac{1-3w}{3(1+w)}\ln
\left( \frac{\rho_{reh}^{1/4}}{\rho_{end}^{1/4}}\right)+\ln \left(
\frac{\rho_k^{1/4}}{10^{16}\mathrm{GeV}}\right)\, ,
\end{equation}
where $\rho_k$ is the energy density of the Universe at the
beginning of the inflationary era, when the mode $k$ exited the
horizon. Also the pivot scale $k_*$ is $k_*=0.05$Mpc$^{-1}$ and in
addition a major assumption we shall take into account in the
following is that effective degrees of freedom of particles during
inflation, until the end of the radiation domination era is
constant. Thus we neglect mechanisms which may effectively change
the effective degrees of freedom during reheating, for example
supersymmetry breaking or other mechanisms. In principle, the
neglected cases can also be dealt in a similar way to that we
shall demonstrate, with slight changes incorporated in the
effective degrees of freedom parameter $g_*$ which enters in the
general relation for all energy densities and the corresponding
temperature $\rho=\frac{\pi^2}{30}g_*T^4$. With this assumption,
we can easily rewrite Eq. (\ref{efoldingsmainrelation}) in terms
of the corresponding temperatures instead of the energy densities.

Let us now proceed to quantify our findings in terms of several
possible scenarios for the constant $w$ short post-inflationary
era. As we mentioned, our main assumption is that this short
constant $w$ era is generated by $f(R)$ gravity, thus the geometry
effectively generates this era. In general, this short era can
have any value in the range $-\frac{1}{3} \leq w \leq 1$. For the
purposes of this work we shall confine ourselves in two scenarios
listed below,
\begin{align}\label{scenarios}
&\mathrm{Scenario}\,\,I\,:\,\,\mathrm{EoS}\,\,w=-\frac{1}{3}\,\,(\ddot{a}=0)\,
,\\ \notag &
\mathrm{Scenario}\,\,II\,:\,\,\mathrm{EoS}\,\,w=0\,\,(\mathrm{Early}\,\,\,\mathrm{Matter}\,\,\,\mathrm{Domination})\,
.
\end{align}
The Scenario I corresponds to a short-lasting post-inflationary
era with EoS parameter equal to $w=-1/3$, for which the Universe
neither accelerates nor decelerates, since $\ddot{a}=0$ for
$w=-1/3$. This evolution is known as thermal inflation in the
literature \cite{Burgess:2005sb}. The Scenario II corresponds to
an early matter domination era with EoS parameter $w=0$. In both
cases, these short post-inflationary eras have geometric origin
and are generated by the $f(R)$ gravity $F_w(R)$, both in the
presence of matter and radiation perfect fluids. The perfect
fluids play an also important role, since post-inflationary these
cannot be neglected, as in the inflationary case. Thus the synergy
of $f(R)$ gravity and matter and radiation perfect fluids gives
rise to these $w$-EoS post-inflationary eras. Finally, we shall
consider three different reheating temperatures, listed below,
\begin{align}\label{scenarios}
&
\mathrm{High}\,\,\,\mathrm{Reheating}\,\,\,\mathrm{Temperature}:\,\,\,T_R=10^{12}\mathrm{GeV}\, ,\\
\notag &
\mathrm{Intermediate}\,\,\,\mathrm{Reheating}\,\,\,\mathrm{Temperature}:\,\,\,T_R=10^{7}\mathrm{GeV}\, ,\\
\notag &
\mathrm{Low}\,\,\,\mathrm{Reheating}\,\,\,\mathrm{Temperature}:\,\,\,T_R=10^{2}\mathrm{GeV}\,
.
\end{align}
In the literature there exist scenarios which motivate the
low-reheating temperature case, in some cases even motivate MeV
scale reheating temperatures \cite{Hasegawa:2019jsa}.
\begin{table}[h!]
  \begin{center}
    \caption{\emph{\textbf{Scenario I: EoS $w=-\frac{1}{3}$ ($\ddot{a}$)}}}
    \label{table1}
    \begin{tabular}{|r|r|r|r|}
     \hline
      \textbf{Observational Indices and $e$-foldings} & \textbf{$T_R=10^{12}$GeV} & \textbf{$T_R=10^{7}$GeV} & \textbf{$T_R=10^{2}$GeV}\\
           \hline
           $e$-foldings number $N$ & 63.7413 & 52.2284& 40.7155\\ \hline
      Tensor-to-Scalar Ratio $r$ & 0.00295352 & 0.00439914 & 0.00723873\\ \hline
      Tensor Spectral Index $n_T$ & -0.0000615316 & -0.0000916488 & -0.000150807 \\ \hline
    \end{tabular}
  \end{center}
\end{table}
Let us now proceed to the evaluation of the observational indices
relevant to the calculation of the energy spectrum of the
primordial gravitational waves. Specifically, we shall calculate
the tensor-to-scalar ratio and the tensor spectral index, for the
Scenarios I and II, and for the three distinct reheating
temperatures. We shall be interested in modes with
$k=0.05$Mpc$^{-1}$, which is the pivot scale used in Planck. For
$f(R)$ gravity, the tensor-to-scalar ratio is
\cite{reviews1,Odintsov:2020thl,Odintsov:2021kup},
\begin{equation}\label{tensotoscalarratio}
r=48\epsilon_1^2\, ,
\end{equation}
and the tensor spectral index for $f(R)$ gravity is
\cite{reviews1,Odintsov:2020thl,Odintsov:2021kup},
\begin{equation}\label{tensorspectralindexr2ini}
n_T\simeq -2\epsilon_1^2\, ,
\end{equation}
where $\epsilon_1$ is the first slow-roll index
$\epsilon_1=-\dot{H}/H^2$. Since the inflationary era is
controlled by an $R^2$ gravity in the present context, the first
slow-roll index is $\epsilon_1\simeq \frac{1}{2N}$, hence the
resulting tensor spectral index is,
\begin{equation}\label{r2modeltensorspectralindexfinal}
n_T\simeq -\frac{1}{2N^2}\, ,
\end{equation}
while the tensor-to-scalar ratio for the $R^2$ model is,
\begin{equation}\label{tensortoscalarfinal}
r=\frac{12}{N^2}\, .
\end{equation}
Now the $e$-foldings number is affected by both the reheating
temperature and the EoS parameter $w$, as is dictated by relation
(\ref{efoldingsmainrelation}). Hence we will calculate the
$e$-foldings number for the two scenarios of Eq. (\ref{scenarios})
and for the three different reheating temperatures
(\ref{scenarios}). The results for the Scenarios I and II can be
found in Tables \ref{table1} and \ref{table2}.
\begin{table}[h!]
  \begin{center}
    \caption{\emph{\textbf{Scenario II: EoS $w=0$ (Early Matter Domination)}}}
    \label{table2}
    \begin{tabular}{|r|r|r|r|}
     \hline
      \textbf{Observational Indices and $e$-foldings} & \textbf{$T_R=10^{12}$GeV} & \textbf{$T_R=10^{7}$GeV} & \textbf{$T_R=10^{2}$GeV}\\
           \hline
           $e$-foldings number $N$ & 65.3439 & 61.5063& 57.6687\\ \hline
      Tensor-to-Scalar Ratio $r$ & 0.00281042 & 0.00317206 & 0.00360829\\ \hline
      Tensor Spectral Index $n_T$ & -0.0000585503 & -0.0000660847 & -0.0000751727 \\ \hline
    \end{tabular}
  \end{center}
\end{table}
For both Scenarios I and II, there exist several mentionable
features, for example in both Scenarios I and II for the large
reheating temperature case, the $e$-foldings number exceeds
$N=60$. Also for the Scenario I, the low reheating temperature
case results to a small $e$-foldings number. Finally, for Scenario
II, all the $e$-foldings numbers satisfy $N>50$. In the next
section we shall use the tensor-to-scalar ratio and the tensor
spectral index from Tables \ref{table1} and \ref{table2}, in order
to calculate the predicted energy spectrum of the primordial
gravitational waves for the $f(R)$ gravity theoretical framework
developed in this paper.

\section{Primordial Gravitational Wave Energy Spectrum Amplification Due to non-canonical $f(R)$ Reheating}

In the literature there exist several works on theoretical
predictions of the energy spectrum of the primordial gravitational
waves
\cite{Kamionkowski:2015yta,Denissenya:2018mqs,Turner:1993vb,Boyle:2005se,Zhang:2005nw,Schutz:2010xm,Sathyaprakash:2009xs,Caprini:2018mtu,
Arutyunov:2016kve,Kuroyanagi:2008ye,Clarke:2020bil,Kuroyanagi:2014nba,Nakayama:2009ce,Smith:2005mm,Giovannini:2008tm,
Liu:2015psa,Zhao:2013bba,Vagnozzi:2020gtf,Watanabe:2006qe,Kamionkowski:1993fg,Giare:2020vss,Kuroyanagi:2020sfw,Zhao:2006mm,
Nishizawa:2017nef,Arai:2017hxj,Bellini:2014fua,Nunes:2018zot,DAgostino:2019hvh,Mitra:2020vzq,Kuroyanagi:2011fy,Campeti:2020xwn,
Nishizawa:2014zra,Zhao:2006eb,Cheng:2021nyo,Nishizawa:2011eq,Chongchitnan:2006pe,Lasky:2015lej,Guzzetti:2016mkm,Ben-Dayan:2019gll,
Nakayama:2008wy,Capozziello:2017vdi,Capozziello:2008fn,Capozziello:2008rq,Cai:2021uup,Cai:2018dig,Odintsov:2021kup,Benetti:2021uea,Lin:2021vwc,Zhang:2021vak,Odintsov:2021urx,Pritchard:2004qp,Zhang:2005nv,Baskaran:2006qs,Oikonomou:2022xoq,Odintsov:2022cbm,Odintsov:2022sdk}
and in about fifteen years from now, the experiments will start to
yield data. Thus very soon from now, several theoretical
frameworks will be put to test. Most of the experiments will probe
modes that re-entered the Hubble horizon during the dark era of
reheating and during the radiation domination era. In this work we
shall be interested in calculating the energy spectrum of the
primordial gravitational waves, generated by an underlying $f(R)$
gravity theory which we developed in the previous section. Our
main interest is to see the synergistic effect of a short $f(R)$
gravity generated era with constant EoS parameter $w$, in the
presence of matter and radiation fluids.

Let us fix the duration of this short $w$-era, and without loss of
generality, let us assume that it lasts from the end of inflation,
so from temperature $T_{end}\sim 10^{13}$GeV, until
$T_{pr}=10^{12}$GeV. Note that the high scale reheating case we
discussed in the previous section, corresponds to the same
temperature as $T_{pr}=10^{12}$GeV. In order to calculate the
exact effect of the constant $w$ $f(R)$ gravity generated era, we
need to find the redshifts corresponding to the temperatures
$T_{end}$ and $T_{pr}$. Using the relation $T=T_0(1+z)$
\cite{Garcia-Bellido:1999qrp}, with $T_0$ being the present day
temperature $T_0=2.58651\times 10^{-4}$eV, we easily find that the
redshift corresponding to $T_{end}$ is $z_{end}=3.86621\times
10^{16}$ while the redshift corresponding to $T_{pr}$ is
$z_{pr}=3.86621\times 10^{15}$.
\begin{figure}[h!]
\centering
\includegraphics[width=40pc]{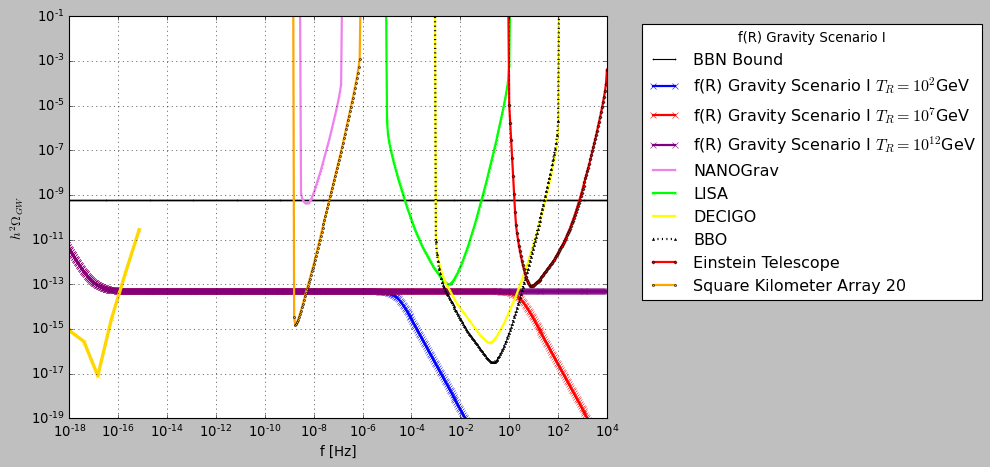}
\caption{The $h^2$-scaled gravitational wave energy spectrum for
the $w$-era $f(R)$ gravity, and for the Scenario I. The $f(R)$
gravity curves correspond to $T_R=10^{12}$GeV (purple), to
$T_R=10^{7}$GeV (red) and to (blue) $T_R=10^{2}$GeV.}
\label{plotfinalfrpure1}
\end{figure}
Now let us briefly recall how to calculate the overall effect of
$f(R)$ gravity on the energy spectrum of the primordial
gravitational waves, from present day at redshift $z=0$ up to the
inflationary era. Details on this analysis can be found
\cite{Odintsov:2021kup}. The parameter that quantifies the overall
effect of $f(R)$ gravity on the general relativistic waveform is
$\alpha_M$, defined as,
\begin{equation}\label{amfrgravity}
a_M=\frac{f_{RR}\dot{R}}{f_RH}\, ,
\end{equation}
and the deformation of the general relativistic waveform is
\cite{Nishizawa:2017nef,Arai:2017hxj},
\begin{equation}\label{mainsolutionwkb}
h=e^{-\mathcal{D}}h_{GR}\, ,
\end{equation}
with $h_{GR}$ being the general relativistic waveform with
$a_M=0$, and $\mathcal{D}$ being defined as,
\begin{equation}\label{dform}
\mathcal{D}=\frac{1}{2}\int^{\tau}a_M\mathcal{H}{\rm
d}\tau_1=\frac{1}{2}\int_0^z\frac{a_M}{1+z'}{\rm d z'}\, .
\end{equation}
Let us here quote in brief some details regarding the derivation
of Eqs. (\ref{dform})-(\ref{amfrgravity}). For more details we
refer the reader to Ref. \cite{Odintsov:2021kup} and more
importantly to Ref. \cite{Odintsov:2022cbm}. The differential
equation which is obeyed by the Fourier transformation of the
primordial tensor perturbation $h_{i j}$, has the following form,
\begin{equation}\label{mainevolutiondiffeqnfrgravity}
\ddot{h}(k)+\left(3+a_M \right)H\dot{h}(k)+\frac{k^2}{a^2}h(k)=0\,
,
\end{equation}
where $\alpha_M$ is defined as follows,
\begin{equation}\label{amfrgravity}
a_M=\frac{\dot{Q}_t}{Q_tH}\, ,
\end{equation}
and the function $Q_t$ is unique for every distinct modified
gravity. For a complete list of all the functional forms of $Q _t$
and $a_M$ we refer the reader to the review
\cite{Odintsov:2022cbm}. Basically, the parameter $a_M$ quantifies
the overall effect of the modified gravity on the tensor
perturbations evolution. For a general $f(R,\phi)$ gravity with
action,
\begin{equation}\label{action1}
\mathcal{S}=\int
\mathrm{d}^4x\sqrt{-g}\Big{(}\frac{f(R,\phi)}{2}-\frac{\omega(\phi)}{2}\partial^{\mu}\phi\partial_{\mu}\phi-V(\phi)\Big{)}\,
,
\end{equation}
the parameter $Q_t$ is $Q_t=\frac{1}{\kappa^2}\frac{\partial
f(R,\phi)}{\partial R}$, where $\kappa=\frac{1}{M_p}$, where $M_p$
is the reduced Planck mass. Therefore, for a pure $f(R,\phi)$
gravity, the parameter $a_M$ has the following form,
\begin{equation}\label{amfrphi}
a_M=\frac{\frac{\partial^2f}{\partial R \partial
\phi}\dot{\phi}+\frac{\partial^2f}{\partial
R^2}\dot{R}}{\frac{\partial f}{\partial R}H}\, ,
\end{equation}
which reduces to that of Eq. (\ref{amfrgravity}) when a pure
$f(R)$ gravity is considered, in which case $f(R,\phi)=f(R)$.
Using the conformal time $\tau$, the evolution equation
(\ref{mainevolutiondiffeqnfrgravity}) is written as,
\begin{equation}\label{mainevolutiondiffeqnfrgravityconftime}
h''(k)+\left(2+a_M \right)\mathcal{H} h'(k)+k^2h(k)=0\, ,
\end{equation}
where the ``prime'' indicates differentiation with respect to the
conformal time $\tau$, and furthermore we also defined
$\mathcal{H}=\frac{a'}{a}$. The WKB solution of the above
differential equation shall now be extracted, considering only
subhorizon modes which satisfy Eq.
(\ref{mainevolutiondiffeqnfrgravityconftime}). Recall that the
subhorizon modes will directly be probed by the future
gravitational wave experiments, since these modes became
subhorizon modes during the reheating era. Lets consider a WKB
solution of the following form
$h_{ij}=\mathcal{A}e^{i\mathcal{B}}h_{ij}^{GR}$, describes
theories with gravitational wave speed equal to unity in natural
units, so in this case the full WKB solution of Eq.
(\ref{mainevolutiondiffeqnfrgravityconftime}) is
\cite{Nishizawa:2017nef,Arai:2017hxj},
\begin{equation}\label{mainsolutionwkb}
h=e^{-\mathcal{D}}h_{GR}\, ,
\end{equation}
with $h_{i j}=h e_{i j}$, and with $h_{GR}$ denoting the general
relativistic waveform which note that it is the solution of the
differential equation
(\ref{mainevolutiondiffeqnfrgravityconftime}) with $a_M=0$. After
discussing these important issues, let us proceed to the
gravitational wave energy spectrum.

\begin{figure}[h!]
\centering
\includegraphics[width=40pc]{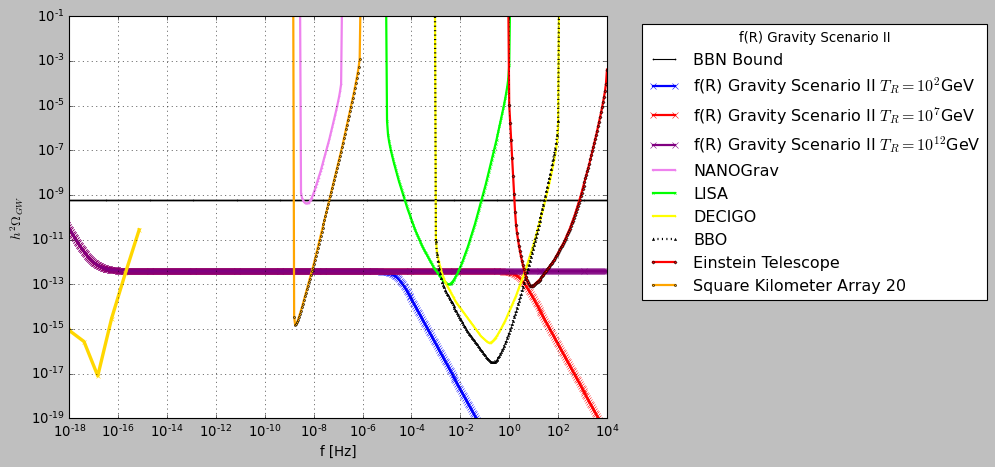}
\caption{The $h^2$-scaled gravitational wave energy spectrum for
the $w$-era $f(R)$ gravity, and for the Scenario II. The $f(R)$
gravity curves correspond to $T_R=10^{12}$GeV (purple), to
$T_R=10^{7}$GeV (red) and to (blue) $T_R=10^{2}$GeV.}
\label{plotfinalfrpure2}
\end{figure}

The energy spectrum of the primordial gravitational waves
including the effects of $f(R)$ gravity is
\cite{Boyle:2005se,Nishizawa:2017nef,Arai:2017hxj,Nunes:2018zot,Liu:2015psa,Zhao:2013bba,Odintsov:2021kup},
\begin{align}
\label{GWspecfR}
    &\Omega_{\rm gw}(f)=e^{-2\mathcal{D}}\times \frac{k^2}{12H_0^2}r\mathcal{P}_{\zeta}(k_{ref})\left(\frac{k}{k_{ref}}
\right)^{n_T} \left ( \frac{\Omega_m}{\Omega_\Lambda} \right )^2
    \left ( \frac{g_*(T_{\rm in})}{g_{*0}} \right )
    \left ( \frac{g_{*s0}}{g_{*s}(T_{\rm in})} \right )^{4/3} \nonumber  \left (\overline{ \frac{3j_1(k\tau_0)}{k\tau_0} } \right )^2
    T_1^2\left ( x_{\rm eq} \right )
    T_2^2\left ( x_R \right )\, ,
\end{align}
where $k_{ref}=0.002$$\,$Mpc$^{-1}$ is the CMB pivot scale, $n_T$
is the tensor spectral index and $r$ is the tensor-to-scalar
ratio. Thus, the vital element of the evaluation of the energy
spectrum is calculating the parameter $\mathcal{D}$. For the
evaluation, we need to divide the integration periods to two, with
the first being from present day up to redshift
$z_{pr}=3.86621\times 10^{15}$ and the second from the redshift
$z_{pr}$ up to $z_{end}=3.86621\times 10^{16}$ exactly at the end
of the inflationary era. In these two intervals, the dominant
$f(R)$ gravity form is different, so the $a_M$ parameter defined
in (\ref{amfrgravity}) is different, and specifically we have,
\begin{equation}\label{dformexplicitcalculation}
\mathcal{D}=\frac{1}{2}\left(\int_0^{z_{pr}}\frac{a_{M_1}}{1+z'}{\rm
d z'}+\int_{z_{pr}}^{z_{end}}\frac{a_{M_2}}{1+z'}{\rm d
z'}\right)\, ,
\end{equation}
where the parameters $a_{M_1}$ and $a_{M_2}$ being evaluated for
$f(R)\sim R+f_{DE}(R)$ and $f(R)\sim F_w(R)$ respectively. For the
first integral, as it was shown \cite{Odintsov:2021kup}, the
contribution is almost zero, so this can be neglected. For the
second integral, for the high reheating temperature case, the
contribution for $w=-1/3$ is
$\int_{z_{pr}}^{z_{end}}\frac{a_{M_2}}{1+z'}{\rm d z'}=-12.434$
and for $w=0$ the contribution is
$\int_{z_{pr}}^{z_{end}}\frac{a_{M_2}}{1+z'}{\rm d z'}=-14.5063$.
Hence the overall amplification factors are of the order
$\mathcal{O}(10^{5})$  and $\mathcal{O}(10^{6})$ respectively. We
evaluated the energy spectrum for both the Scenario I ($w=-1/3$)
and II ($w=0$) and for all the reheating temperatures, and the
results of our analysis appear in Figs. \ref{plotfinalfrpure1} and
\ref{plotfinalfrpure2}, where we plot the $f(R)$ gravity
$h^2$-scaled energy spectrum versus the frequency. In the plots
the sensitivity curves of most of the future gravitational waves
experiments are shown too. We used the three different reheating
temperatures and specifically, the purple curve corresponds to the
high reheating temperature $T_R=10^{12}$GeV, the red curve
corresponds to the intermediate reheating temperature
$T_R=10^{7}$GeV and the blue curve corresponds to the low
reheating temperature $T_R=10^{2}$GeV.
As it can be seen in Figs. \ref{plotfinalfrpure1} and
\ref{plotfinalfrpure2}, in both cases for Scenarios I and II, the
low-reheating temperature signal will be undetectable from some
future experiments, and also the intermediate reheating
temperature case will be undetectable by the Einstein telescope.
Finally let us note that the signal produced by the Scenario II is
larger compared to that of Scenario I.

In conclusion, the results indicate that even a short period with
EoS parameters $w=-1/3$ and $w=0$ generated by $f(R)$ gravity
synergistically with matter and radiation fluids, will be
detectable by future gravitational wave experiments. This result
changes our perspective on primordial gravitational waves
experiments, since it shows that even $f(R)$ gravity may lead to
detectable primordial gravitational waves signals.

Finally, let us briefly discuss the scale of the reheating
temperature. Basically, this era is known in cosmology as the dark
age, since we know nothing about the reheating and the subsequent
radiation domination era. In contrast to the post-recombination
era for $z>1100$ which we know very well via the CMB, to date we
have only hints that the reheating era existed, since after
inflation the Universe was too cold and some mechanism should heat
up the Universe in order for it to proceed to the formation of
galaxies. This is the reheating mechanism after inflation, but we
do not know the essential features of this era and the subsequent
radiation domination era. We know that the reheating temperature
should be smaller than the low scale inflation scenario
temperature, so smaller than $T\leq 10^{12}\,$GeV, but nothing
dictates how smaller it should be. In fact, there are scenarios
which indicate that the reheating temperature could be of the MeV
scale order, see for example \cite{Hasegawa:2019jsa}. Hopefully,
the gravitational wave interferometer experiments will also shed
some light on this issue too.

\section{Conclusions}

In this paper we presented a mechanism which can cause a
measurable amplification in the energy spectrum of primordial
gravitational waves of $f(R)$ gravity theories. The latter are
known to produce an undetectable energy spectrum, however, as we
showed, if post-inflationary the $f(R)$ gravity synergistically
with matter and radiation fluids generate a short period with
constant EoS parameter, this era can cause a significant
amplification of the energy spectrum of primordial gravitational
waves. We considered two cases of interest, one with EoS parameter
$w=-1/3$ in which case the Universe neither accelerates nor
decelerates, and also one case with $w=0$, so a primordially early
matter domination era. Both these eras are realized by the
synergistic action of $f(R)$ gravity and of the perfect matter
fluids present. This is in contrast with the general relativistic
effects of a constant EoS parameter, in which case the
amplification is minor. For the purposes of our analysis, we
assumed that during inflation, the dominant part of the
inflationary Lagrangian is controlled by an $R^2$ gravity, and for
the inflationary era the perfect matter fluids do not affect the
evolution. After the inflationary era, and for a short period, the
evolution was assumed to be a power-law type as a function of the
cosmic time, with constant EoS parameter. This era we assumed that
it was generated by $f(R)$ gravity and the perfect fluids. After
this short constant EoS era, the radiation and matter domination
eras occurred controlled by the corresponding matter fluids,
followed by the late-time dark energy era. For the inflationary
era we also included the effects of the post-inflationary era on
the $e$-foldings number of inflation. An interesting extension and
generalization of this work is to consider the effects of an early
dark energy era on the present day evaluated energy spectrum of
the primordial gravitational waves for $f(R)$ gravity. With early
dark energy we mean for redshifts near $z\sim 3400$ near the
matter radiation equality. Another interesting issue that should
be addressed is the comparison of the energy spectrum of
primordial gravitational waves generated by a geometrically
generated post-inflationary era, with the general relativistic
energy spectrum. In both cases, the modes that re-enter the
horizon during the constant EoS parameter era, will be affected,
and the crucial issue is to compare in detail the $f(R)$ and
general relativistic results. We hope to address these issues in
the future.

The future in gravitational wave astronomy will be particularly
interesting both theoretically and experimentally especially if
signals of a stochastic gravitational background is detected by
all, or even by some gravitational wave experiments. Let us
discuss in brief these issues and also we shall try to speculate
on the possible outcomes of the gravitational wave experiments.
Specifically, if no signal of primordial gravitational waves is
detected by all the experiments, would that be bad news for
inflation? The answer is no, because the sensitivities of future
interferometers like LISA, DECIGO and BBO might be higher than the
$h^2$-scaled gravitational waves energy spectrum of inflation.
Thus no model of inflation can be directly excluded by
observations, it is possible that simple scenarios like scalar
field inflation or $f(R)$ gravity inflation might drive the
inflationary dynamics. In such a scenario, one is certain that the
tensor spectral index is not positive, since such a feature would
lead to a detection of the stochastic gravitational wave signal.
However, in such a scenario it would be hard to determine or have
hints about the scale of the reheating temperature. Now let us
assume that a signal is observed by all future gravitational wave
experiments. This would be a sensational result because this would
exclude simple inflationary scenarios, like scalar field inflation
and $f(R)$ gravity, without any abnormal reheating era, like the
one we considered in this paper. Thus, the results of this paper
apply to the detection of a signal, which they can explain too.
But the plot thickens since very careful combination of all the
results and for a wide frequency range must be performed in order
to discriminate different scenarios which can lead to a detection
of s signal. The form of the signal can yield hints on the
reheating temperature, on the particle content and the EoS of the
Universe when the tensor modes reentered the horizon. Also if a
signal is detected by some experiments, and not by others, may
strongly favor supersymmetry breaking for the frequencies range
for which the stochastic gravitational wave signal is absent.
Hence if a signal is observed in future gravitational wave
experiments, the questions are, is this due to a large reheating
temperature with a positive tensor spectral index or due to an
abnormal geometrically generated reheating era? This is not easy
to answer, so we anticipate all the observational data from all
the future gravitational wave experiments to further analyze the
results.

\end{document}